\documentclass[sigconf]{acmart}

\usepackage{enumitem}
\usepackage{bbding}
\usepackage{graphicx} 
\AtBeginDocument{%
  \providecommand\BibTeX{{%
    \normalfont B\kern-0.5em{\scshape i\kern-0.25em b}\kern-0.8em\TeX}}}

\setcopyright{acmcopyright}
\copyrightyear{2024}
\acmYear{2024}
\acmDOI{}

\acmConference[Q-SE 2024]{5th International Workshop on Quantum Software Engineering}{April 2024}{Lisbon, Portugal}
\begin{document}

\title{On Repairing Quantum Programs Using ChatGPT}

\author{Xiaoyu Guo}
\affiliation{%
  \institution{Kyushu University}
  \country{Fukuoka, Japan}
}
\email{guo.xiaoyu.961@s.kyushu-u.ac.jp}

\author{Jianjun Zhao}
\authornote{Corresponding author}
\affiliation{%
  \institution{Kyushu University}
  \country{Fukuoka, Japan}
}
\email{zhao@ait.kyushu-u.ac.jp}

\author{Pengzhan Zhao}
\affiliation{%
  \institution{Kyuhu University}
  \country{Fukuoka, Japan}
}
\email{zpz2393247079@gmail.com}



\begin{abstract}
Automated Program Repair (APR) is a vital area in software engineering aimed at generating automatic patches for vulnerable programs. While numerous techniques have been proposed for repairing classical programs, the realm of quantum programming lacks a comparable automated repair technique. In this initial exploration, we investigate the use of ChatGPT for quantum program repair and evaluate its performance on Bugs4Q, a benchmark suite of quantum program bugs.
Our findings demonstrate the feasibility of employing ChatGPT for quantum program repair. Specifically, we assess ChatGPT's ability to address bugs within the Bugs4Q benchmark, revealing its success in repairing 29 out of 38 bugs. This research represents a promising step towards automating the repair process for quantum programs.
\end{abstract}



\maketitle

\section{Introduction}
\label{sec:introduction}

Quantum computing promises an exponential increase in computational power, revolutionizing fields such as cryptography, optimization, and materials science~\cite{barends2014superconducting,guerreschi2017practical,mosca2018cybersecurity,grimsley2019adaptive}. However, along with this transformative potential comes the challenge of programming these complex quantum systems, which has become increasingly complex with the expansion of quantum hardware and software and has raised concerns about introducing programming errors.
Quantum computing is undeniably significant in our technological landscape. It has opened the door to solving previously unsolvable problems and achieved breakthroughs in critical areas~\cite {olson2017quantum}. However, these promises can only be realized if quantum programs exhibit robustness, reliability, and absence of errors.

Automated Program Repair (APR)~\cite{le2011genprog,gazzola2018automatic,le2021automatic} plays a vital role in addressing software bugs in classical computing. It simplifies debugging, improves software quality, and shortens development cycles by automatically generating patches for vulnerable programs. However, as we step into the emerging field of quantum computing, a glaring gap is evident - the lack of automatic repair techniques tailored for quantum programs. While numerous methods have been proposed for repairing classical programs using traditional APR techniques~\cite{le2011genprog,gazzola2018automatic,le2021automatic} and Large Language Models (LLMs) such as ChatGPT~\cite{xia2023keep,tian2023chatgpt}, the realm of quantum programming lacks a comparable automated repair technique.

Quantum programming presents several unique challenges that are distinct from classical software development. Quantum programs operate according to the principles of quantum mechanics, governed by phenomena such as superposition and quantum interference, and are fundamentally different from classical computing. This quantum environment generates unique types of bugs~\cite{huang2019statistical,zhao2023bugs4q,zhao2021bugs4q-a,zhao2021identifying,campos2021qbugs,paltenghi2022bugs}, including the bugs related to the quantum gate, quantum measurement, quantum entanglement, and quantum noise, which are often hidden and difficult to find and fix.
Moreover, the primary quantum computing hardware in use is the Noisy Intermediate-Scale Quantum (NISQ) device~\cite{preskill2018quantum}, known for its susceptibility to errors and limited resources. These hardware limitations make quantum programming more complex and highlight the need for robust bug repair techniques.
However, the challenge is that we currently lack specialized repair techniques for quantum programs~\cite{zhao2020quantum}. Traditional bug-fixing methods, designed for classical software, struggle when applied in the unique quantum computing environment. Quantum-specific knowledge is crucial. Hence, there is an urgent requirement for innovative quantum-specific repair techniques to simplify the process of fixing bugs in quantum programs."



In this paper, we address this gap by investigating the use of ChatGPT~\cite{chatgpt2022} for quantum program repair and evaluating its performance on Bugs4Q, a benchmark suite of quantum program bugs. Our findings demonstrate the feasibility of employing ChatGPT for quantum program repair. Specifically, we assess ChatGPT's ability to address bugs within the Bugs4Q benchmark, revealing its success in repairing 29 out of 38 bugs. This research represents a promising step towards automating the repair process for quantum programs.

In summary, this paper makes the following contributions:
\begin{itemize}[leftmargin=2em]
    \item We demonstrate ChatGPT's effective capability in repairing quantum programs.
    \item We evaluate ChatGPT's performance in program repair using a real-world benchmark, examining various dialogue function settings.
     \item We discuss the challenges of ChatGPT as a semi-automated program repair compared to manual repair when applied to quantum program repair. The discussion contributes actionable insights to adopting LLMs for quantum program repair and offers a better understanding of ChatGPT’s practical applications in the software engineering community.
\end{itemize}

The rest of this paper is structured as follows: In Section~\ref{sec:background}, we provide background information, including quantum computing, large language models, and automatic program repair. Section~\ref{sec:methodology} outlines our methods. We present our experimental design and results in Section~\ref{sec:experiments}. Section~\ref{sec:discussion} is dedicated to discussions. We delve into related work in Section~\ref{sec:related-work}. Finally, we conclude our work in Section~\ref{sec:conclusion}.

\section{Background}
\label{sec:background}
We next introduce some basics of quantum computing, followed by a brief overview of the concepts of APR in classical software and the introduction of ChatGPT. 

\subsection{Quantum Computing}

\subsubsection{Basic Properties of Qubits}
A quantum bit, or qubit, represents the quantum analog of a classical bit but possesses distinct properties. Unlike a classical bit, which behaves like a coin with just two states, 0 and 1, a qubit can exist in a continuum of states between $|0\rangle$ and $|1\rangle$. Mathematically, we represent a qubit as:

\begin{equation}
|\psi\rangle = \alpha|0\rangle + \beta|1\rangle,\ \text{where}\ |\alpha|^2 + |\beta|^2 = 1
\end{equation}

Until measured, we cannot definitively ascertain whether the qubit is in state $|0\rangle$ or $|1\rangle$. Here, $\alpha$ and $\beta$ denote the amplitudes of the qubit, with measurement probabilities corresponding to $|\alpha|^2$ and $|\beta|^2$. Upon measurement, the probability of observing $|0\rangle$ is $|\alpha|^2$, while the probability of observing $|1\rangle$ is $|\beta|^2$.

In quantum programming, operations are represented by quantum logic gates. These gates perform various functions, including rotating qubits to arbitrary angles, introducing superposition to qubits, and establishing controlled connections between qubits. Some gates apply to individual qubits, while others operate on multiple qubits.

Quantum circuits consist of interconnected quantum gates, with qubits, gates, and measurements organized sequentially. Quantum programs are combinations of quantum circuits and classical programs. In this work, we consider quantum programs as either quantum circuits or combinations of quantum circuits, recognizing that the definition of quantum programs may vary.

\subsubsection{Qiskit Programming Framework}
\label{subsec:qiskit}

In recent years, several open-source programming frameworks have emerged to facilitate quantum programming, empowering the implementation and utilization of quantum algorithms. Noteworthy among these frameworks are Qiskit~\cite{ibm2021qiskit}, Q\#~\cite{svore2018q}, Cirq~\cite{cirq2018google}, 
isQ~\cite{Guo2022isQ}, and ProjectQ~\cite{haner2016high}. These frameworks serve as essential tools for quantum program development. For our study, we have selected Qiskit, a widely adopted quantum programming language, as our primary platform of choice.

Qiskit, a Python package, equips developers with a comprehensive suite of tools for creating and manipulating quantum programs capable of running on both prototype quantum devices and simulators~\cite{aleksandrowicz2019qiskit}. Beyond its fundamental capabilities, Qiskit offers built-in modules for noise characterization and circuit optimization, effectively mitigating the impact of quantum noise on computations. Furthermore, it boasts an extensive library of quantum algorithms tailored for applications in machine learning, optimization, and chemistry.

Figure \ref{figure-3} provides a simplified Qiskit program as an illustrative example. The program commences by initializing quantum and classical registers, followed by the registration of two qubits, each initialized to the state $|0\rangle$. Subsequently, an \textit{H} gate is applied to the first qubit, entangling it into the state $\frac{1}{\sqrt{2}}|0\rangle + \frac{1}{\sqrt{2}}|1\rangle$, while an \textit{X} gate flips the second qubit from $|0\rangle$ to $|1\rangle$. Finally, a measurement operation is performed, with the outcomes recorded in a classical array.

\begin{figure}[h]	
\centering
\resizebox{0.38\textwidth}{!}{
\begin{tabular}{@{}l@{}}
\toprule
\ 1 \quad   from qiskit import * \\    
\ 2 \quad   qubit = QuantumRegister(2)\\  
\ 3 \quad   bit = ClassicalRegister(2)\\
\ 4 \quad   prog = QuantumCircuit(qubit, bit)\\
\ 5 \quad   prog.h(0)\\
\ 6 \quad   prog.x(1)\\
\ 7 \quad   for i in range(2):\\
\ 8 \quad   \hspace{0.5cm} prog.measure(qubit[i], bit[i])\\
\ 9 \quad   backend = Aer.get\_backend('qasm\_simulator')\\
10\quad    job = execute(prog, backend ,shots=1024)\\
11\quad   result = job.result()\\
12\quad   counts = result.get\_counts()  \\
13\quad   print(counts)\\
\bottomrule
\end{tabular}
}
\caption{A simple Qiskit program.}
\label{figure-3}
\end{figure}

\subsection{Automated Classical Program Repair}

Automated Program Repair (APR)~\cite{le2011genprog,gazzola2018automatic,le2021automatic} is a well-established field in classical software engineering that addresses the critical task of automatically generating patches or fixes for software bugs and errors. \textcolor{black}{The concept of APR revolves around the idea of leveraging automated techniques, often rooted in AI and machine learning, to expedite the debugging process.} In classical software development, APR is a valuable tool to enhance software quality, reduce development cycle time, and mitigate the challenges associated with manual debugging.

APR typically involves program synthesis, fault localization, and constraint-solving techniques in classical software. These techniques aim to identify and fix bugs, ranging from logical errors and null pointer exceptions to more complex issues, without human intervention. The effectiveness of APR has been demonstrated across various programming languages and software domains, making it an indispensable component of modern software development practices.


\subsection{Introduction to ChatGPT}

ChatGPT~\cite{chatgpt2022} represents a significant advancement in the field of artificial intelligence and natural language processing. Developed by OpenAI, ChatGPT is a state-of-the-art language model that excels in natural language understanding and generation. It is trained on a vast corpus of text data and has demonstrated remarkable proficiency in comprehending and generating human-readable text.

ChatGPT's capabilities extend to a wide range of natural language tasks, from language translation and text summarization to question answering and content generation. What sets ChatGPT apart is its ability to interpret and generate human language in a manner that is coherent, contextually relevant, and grammatically sound. This natural language interface enables seamless communication with ChatGPT, making it an accessible tool for individuals with diverse levels of expertise. Previous studies~\cite{xia2023keep,sobania2023analysis} have demonstrated that ChatGPT outperforms traditional APR methods and other Large Language Model approaches in program repair due to its capability to comprehend manual descriptions.

In the context of quantum program repair, ChatGPT's natural language understanding and generation capabilities offer a unique opportunity to bridge the gap between quantum program bugs, which are often expressed in plain language by developers, and their corresponding code fixes. By leveraging ChatGPT's language understanding, we can enable quantum developers to describe program bugs in human-readable terms, which are then translated into actionable code fixes, simplifying the bug-fixing process.

\section{Study Methodology}
\label{sec:methodology}

We next present three research questions investigating ChatGPT's ability to fix quantum program bugs. We then present a benchmark and evaluation metrics that illustrate how we have articulated them in response to the research questions.

\subsection{Research Questions}
We evaluate the ChatGPT for repairing quantum program bugs on the following research questions (RQs):

\begin{itemize}[leftmargin=2em]
    \item \textbf{RQ1}: \textit{Can ChatGPT effectively repair bugs in quantum programs?} This question forms the foundation of our work: Can ChatGPT repair quantum programs effectively? The answer to this question determines the feasibility of using LLMs in quantum program repair efforts and sets the stage for subsequent research questions.

    \item \textbf{RQ2}: \textit{What is the performance of ChatGPT when applied to real-world quantum program bugs?} This question centers on assessing the practical utility of ChatGPT. Similar to traditional machine learning models, the efficacy of LLMs must be evaluated through real-world applications.
    
     \item \textbf{RQ3}: \textit{As a semi-automated program repair, how does ChatGPT compare to manual repair?} Given the current absence of fully automated solutions for quantum program repair, programmers often find themselves manually identifying and fixing bugs. This question delves into the challenges of leveraging LLMs to facilitate quantum program repair.
\end{itemize}

We assess ChatGPT's effectiveness in repairing bugs in quantum programs to address RQ1. For RQ2, we employ real-world buggy quantum programs sourced from an available benchmark, Bugs4Q~\cite{zhao2023bugs4q}, as our test subjects. These buggy programs, combined with various hint templates, serve as inputs for ChatGPT. Using the repair rate as a metric, we gauge ChatGPT's performance on real-world bugs. Regarding RQ3, we delve into the practical challenges posed by ChatGPT when employed as a semi-automated program repair tool.

\subsection{Benchmark}
To assess ChatGPT's performance in quantum program repair, we utilize the Bugs4Q benchmark set~\cite{zhao2023bugs4q}, a curated compilation of standard quantum program bug fixes. This benchmark set encompasses a diverse array of quantum program bugs, encompassing logical errors, gate application issues, and quantum measurement problems.

Firstly, Bugs4Q aggregates existing bugs from version control history, along with the real fixes provided by developers. Each real bug is paired with its original and fixed versions, enabling us to assess the correctness of the patches generated by ChatGPT.
Secondly, all bugs within Bugs4Q are reproducible, providing essential data for validating ChatGPT's performance.
Lastly, the variety of bug types in Bugs4Q allows us to comprehensively test ChatGPT's responsiveness to various scenarios.

In this study, we comprehensively retested all current reproducible buggy programs, totaling 38, written in Qiskit from Bugs4Q, ensuring the generalizability of our findings.

\subsection{Evaluation Metrics}
\label{subsec:criteria}

To assess the effectiveness of ChatGPT in repairing quantum program bugs, we define some evaluation metrics as follows:

\begin{itemize}[leftmargin=2em]
\setlength{\itemsep}{2pt}


\item \textit{Correctness}: In evaluating the correctness of the patched code generated by ChatGPT for RQ1 and RQ2, we utilize the provided test cases from the Bugs4Q benchmark. Code that successfully passes all test cases is categorized as correct, whereas any code failing even a single test case is labeled as incorrect. We employ the TOP-5 and AVG-5 metrics as comprehensive performance measures, as elaborated in Section~\ref{subsubsec:randomness}.

\item \textit{Repair rate}: 
We utilize ChatGPT in four different prompt templates. Chat indicates the exclusive use of basic ChatGPT. 
Chat\_D utilizes ChatGPT in conjunction with a basic quantum buggy program description. Chat\_2D builds upon this foundation to offer a more detailed description. Additionally, we have designed Chat\_3D, which directly identifies the location of bugs.
For assessing ChatGPT's performance on these prompt templates, we employ the standard metric of repair rate to further explore beyond correctness. This is calculated as the proportion of patched code generated by ChatGPT that successfully passes all test cases.

\begin{equation}
    Repair\ rate = \frac{correct\ generated\ patches}{all\ generated\ patches}
\end{equation}



\end{itemize}

\subsection{Some Considerations on Using ChatGPT}
We discuss several considerations when using ChatGPT in this study. 
We conduct all our experiments using the ChatGPT web interface version (Released version: Feb 13, 2023, GPT-3.5) and utilize a script to request responses from it automatically.

\subsubsection{The Randomness Nature of ChatGPT}
\label{subsubsec:randomness}

Due to the inherent randomness of ChatGPT responses, ChatGPT may generate different responses to the same requests and prompts, particularly in tasks related to program repair. This variability is because ChatGPT may try different methods to accomplish these repair tasks. To minimize the influence of randomness on the experimental results and ensure their reliability, we systematically tasked ChatGPT with fixing the same quantum program multiple times. Specifically, we initiated five separate conversation requests for the same buggy quantum program, each operating within its independent thread. Within each thread, a maximum of five conversations occurred, representing the number of request-response interactions in the event of an initial failure to fix the bug. We evaluate performance using two measures: (1) TOP-5, which has a value of 1 if at least one of the five attempts of a method solves the program repair problem, and 0 otherwise, and (2) AVG-5, the average success rate of five attempts of a method on the program repair problem. An AVG-5 score of 1 indicates that all five attempts were successful.

\subsubsection{The Prompt Design}
The prompt is critical in utilizing ChatGPT as it provides the context and direction for generating patches. We introduce the prompt we apply for the research experiments. Figure~\ref{figure-1} shows an example of a prompt template that we request ChatGPT for the Quantum program repair task in RQ1, which we follow up on for our experiments. The prompts include asking whether there is a bug in the program (lines 1-2), bug program (lines 4-10), description of the program bug information (line 12), a compiler error message (line 14), current output, and expected output (lines 16-17)

\begin{figure}[t]	
\centering
\resizebox{0.44\textwidth}{!}{
\begin{tabular}{@{}l@{}}
\toprule
\ 1 \quad   \textbf{\textit{D \& 2D:}}  Does the following quantum program have a bug ?  \\    
\ 2 \quad   \textbf{\textit{D \& 2D:}} How to fix it?\\  
\ 3 \quad   \\
\ 4 \quad   from qiskit import *\\
\ 5 \quad   circuit = QuantumCircuit(2)\\
\ 6 \quad   circuit.h(0)\\
\ 7 \quad   circuit.cx(0, 1)\\
\ 8 \quad   result = execute(circuit, backend, shots=1000).result()\\
\ 9 \quad   counts  = result.get\_counts(circuit)\\
10\quad   print(counts)\\
11\quad   \\
12\quad   \textbf{\textit{D \& 2D \&3D:}}\ (Simple Description)  \\
13\quad   How to obtain qubit's amplitude in Qiskit?\\
14\quad   \\
15\quad   \textbf{\textit{2D \& 3D:}}\ (Detailed Description)  \\
16\quad   Is there a way to directly obtain true amplitudes?\\  
17\quad   It's impossible in real life, but I hope the information is     \\
18\quad   contained under the hood of Qiskit for education purposes.\\  
19\quad   \\
20\quad   \textbf{\textit{D \& 2D \& 3D:}}\ (Test Failure Information)\\
21\quad   QiskitError: 'No state vector for experiment "None"'\\
22\quad   \\
23\quad   \textbf{\textit{D \& 2D \& 3D:}}\ (Current Output \& Expect Output) \\
\bottomrule
\end{tabular}
}
\\
\small{D: refer to Chat\_D. 2D: refer to Chat\_2D. 3D: refer to Chat\_3D}
\caption{A prompt example of ChatGPT for program repair.}
\label{figure-1}
\end{figure}

\section{Experiments and Results}
\label{sec:experiments}
We next present the experiments performed to answer our research questions. 

\subsection{Experimental Setup}
Our experimental setup involves the following key steps:

\begin{itemize}[leftmargin=2em]

\item Selection of quantum programs from the Bugs4Q benchmark.
\item Generation of bug descriptions for each selected program, simulating real-world scenarios where quantum developers encounter bugs in their code.
\item Application of ChatGPT to automatically produce code fixes based on the provided bug descriptions.
\item Verification of the generated fixes by running the repaired quantum programs on a quantum simulator or quantum hardware, depending on feasibility and availability.
\item Evaluation of the success of the repair process based on predefined criteria, as outlined in Section~\ref{subsec:criteria}.
\end{itemize}

\subsection{Quantum Program Repair Using ChatGPT}
In this paper, we address this gap by investigating the use of ChatGPT~\cite{chatgpt2022} for quantum program repair and evaluating its performance on Bugs4Q, a benchmark suite of quantum program bugs. Our findings demonstrate the feasibility of employing ChatGPT for quantum program repair. Specifically, we assess ChatGPT's ability to address bugs within the Bugs4Q benchmark, revealing its success in repairing 29 out of 38 bugs. This research represents a promising step towards automating the repair process for quantum programs.

\subsubsection{Experiment Design}
As introduced in Section~\ref{sec:introduction}, we aim to repair 38 buggy codes in the Bugs4Q dataset. We have designed four methods for utilizing ChatGPT in the repair process. The baseline method, Chat, provides only the buggy quantum code without any additional information. We observed that Sobania {\it et al.}~\cite{sobania2023analysis} found that ChatGPT can repair more bugs if provided more information about the bugs through multi-round dialogues. In our experiments, we seek to explore ChatGPT's ability to repair quantum programs by providing additional descriptions of the bugs in the conversation.

 Figure \ref{figure-1} shows an example of Chat\_2D, Chat\_D, and Chat. Each method commences the dialogue with the prompt: "Does this program have a bug? How to fix it?" followed by an empty line and the buggy code without comments. Subsequently, depending on ChatGPT's responses, error information, compiler error messages, or expected input-output is selectively provided. To prevent undue influence on ChatGPT's repair direction by explicitly mentioning the bug in the initial prompt, we devised Chat\_3D, which mirrors Chat\_2D but eliminates the bug inquiry, directly informing ChatGPT that the program contains errors.

We use the TOP-5 metric to evaluate the first five patches generated. If any of these patches pass the test suite, they are considered successful in fixing the buggy code. Therefore, we compared these methods using the TOP-5 metric. In addition, we used the AVG-5 metric, which calculates the probability that one of the first five generated patches passes the test.

\subsubsection{Description Generation} 
\label{subsubsec: Description Generation}
As the essential component of our approach, we generate descriptions in two different forms: a simple description and a detailed one. The simple description briefly describes the symptoms of the bug in the program. Based on a simple description, the detailed version additionally provides the intention of the program and more detailed bug information, such as the location of the misalignment and the bug's behavior. 
For instance, all our descriptions are sourced from bug submitters on GitHub. Here, submitters give descriptions of bugs, summarizing them in the title and explaining the program's purpose and design in the content. We directly extract the descriptions provided by the submitters without any changes because the submitters themselves cannot resolve these buggy programs. If ChatGPT can help repair these buggy programs, which investigate that ChatGPT with human intervention can help developers to solve bugs they solve.

\begin{table}[t]   	
\caption{Results achieved by Chat, Chat\_D, Chat\_2D, and Chat\_3D on the problems from the Bugs4Q benchmark set. We also report the number of successful runs in brackets.}
\label{Table-1}
\resizebox{0.48\textwidth}{!}{%
\begin{tabular}{@{}llcccc@{}}
\toprule
\textbf{Bug ID  }& \textbf{Bug  Type                                  }& \textbf{Chat}                & \textbf{Chat\_D}                & \textbf{Chat\_2D               }& \textbf{Chat\_3D}               \\ \midrule
\#6540  & Parameter Error                            & \XSolidBrush(0/5)                 & \Checkmark(1/5)                 & \Checkmark(2/5)                 & \Checkmark(1/5)                 \\

\#6255  & QASM Calculation                           & \XSolidBrush(0/5)                 & \XSolidBrush(0/5)                 & \XSolidBrush(0/5)                 & \XSolidBrush(0/5)                 \\

\#664   & Measurement Error                          & \XSolidBrush(0/5)                 & \Checkmark(1/5)                 & \Checkmark(5/5)                 & \Checkmark(4/5)                 \\

\#6892  & Parameter Error                            & \XSolidBrush(0/5)                 & \XSolidBrush(0/5)                 & \Checkmark(1/5)                 & \Checkmark(2/5)                 \\

\#5580  & Circuit Design                             & \XSolidBrush(0/5)                 & \XSolidBrush(0/5)                 & \XSolidBrush(0/5)                 & \XSolidBrush(0/5)                 \\

\#5098  & Parameter Error                            & \XSolidBrush(0/5)                 & \Checkmark(3/5)                 & \Checkmark(3/5)                 & \Checkmark(5/5)  \\

\#302   & Parameter Error                            & \Checkmark(5/5)                 & \Checkmark(5/5)                 & \Checkmark(5/5)                 & \Checkmark(5/5)                 \\

\#1337  & Simulator                                  & \XSolidBrush(0/5)                 & \XSolidBrush(0/5)                 & \XSolidBrush(0/5)                 & \XSolidBrush(0/5)                 \\

\#3799  & Output Wrong                               & \XSolidBrush(0/5)                 & \XSolidBrush(0/5)                 & \XSolidBrush(0/5)                 & \XSolidBrush(0/5)                 \\

\#6571  & Function Error                             & \Checkmark(1/5)                 & \Checkmark(5/5)                 & \Checkmark(5/5)                 & \Checkmark(5/5)                 \\

\#135   & Output Wrong                               & \XSolidBrush(0/5)                 & \Checkmark(1/5)                 & \Checkmark(5/5)                 & \Checkmark(5/5)                 \\

\#1119  & Parameter Error                            &\XSolidBrush(0/5)                 & \Checkmark(5/5)                 & \Checkmark(5/5)                 & \Checkmark(5/5)                 \\

\#1446  & Qasm Function                              & \Checkmark(1/5)                 & \Checkmark(5/5)                 & \Checkmark(5/5)                 & \Checkmark(5/5)                 \\

\#1     & Gate  Error                                & \XSolidBrush(0/5)                 & \XSolidBrush(0/5)                 & \Checkmark(3/5)                 & \Checkmark(4/5)                 \\

\#2     & Output Wrong                               & \XSolidBrush(0/5)                 & \Checkmark(1/5)                 & \Checkmark(4/5)                 & \Checkmark(3/5)                 \\

\#3     & Gate Error                                 & \XSolidBrush(0/5)                 & \Checkmark(5/5)                 & \Checkmark(4/5)                 & \Checkmark(5/5)                 \\

\#4     & Gate  Error                                & \Checkmark(1/5)                 & \Checkmark(1/5)                 & \Checkmark(5/5)                 & \Checkmark(5/5)                 \\

\#6     & Gate  Error                                & \XSolidBrush(0/5)                 & \XSolidBrush(0/5)                 & \Checkmark(1/5)                 & \Checkmark(4/5)                 \\

\#9443  & Function Error                             & \Checkmark(1/5)                 & \Checkmark(4/5)                 & \Checkmark(4/5)                 & \Checkmark(5/5)                 \\

\#5557  & Function Error                             & \XSolidBrush(0/5)                 & \Checkmark(3/5)                 & \Checkmark(4/5)                 & \Checkmark(3/5)                 \\

\#8893  & Measurement Error                          & \XSolidBrush(0/5)                 & \Checkmark(5/5)                 & \Checkmark(5/5)                 & \Checkmark(2/5)                 \\

\#7192  & Measurement Error                          & \XSolidBrush(0/5)                 & \Checkmark(5/5)                 & \Checkmark(5/5)                 & \Checkmark(4/5)                 \\

\#20894 & Output Wrong                               & \Checkmark(1/5)                 & \Checkmark(2/5)                 & \Checkmark(4/5)                 & \Checkmark(4/5)                 \\

\#18448 & Gate  Error                                & \XSolidBrush(0/5)                 & \XSolidBrush(0/5)                 & \Checkmark(3/5)                 & \Checkmark(3/5)                 \\

\#15966  & Gate  Error                                & \XSolidBrush(0/5)                 & \Checkmark(5/5)                 & \Checkmark(5/5)                 & \Checkmark(5/5)                 \\

\#9246  & QFE Output Wrong                           & \XSolidBrush(0/5)                 & \XSolidBrush(0/5)                 & \XSolidBrush(0/5)                 & \XSolidBrush(0/5)                 \\

\#5959  & Grover Algrithm                            & \XSolidBrush(0/5)                 & \XSolidBrush(0/5)                 & \XSolidBrush(0/5)                 & \XSolidBrush(0/5)                 \\

\#6692  & Get 2 Measurements from  & \XSolidBrush(0/5)                 & \XSolidBrush(0/5)                 & \XSolidBrush(0/5)                 & \XSolidBrush(0/5)                 \\
  & the Same Execution \\

\#5     & Toffoli Gate Parameter                       & \XSolidBrush(0/5)                 & \XSolidBrush(0/5)                 & \XSolidBrush(0/5)                 & \XSolidBrush(0/5)                 \\

\#7     & QFT Operation                              & \XSolidBrush(0/5)                 & \XSolidBrush(0/5)                 & \Checkmark(1/5)                 & \Checkmark(2/5)                 \\

\#9871  & Custom Four-Qubit Toffoli Gate               & \XSolidBrush(0/5)                 & \XSolidBrush(0/5)                 & \XSolidBrush(0/5)                 & \XSolidBrush(0/5)                 \\

\#8     & Wrong Operation with Gate                  & \XSolidBrush(0/5)                 & \XSolidBrush(0/5)                 & \Checkmark(1/5)                 & \Checkmark(4/5)                 \\

\#17652 & Missing Parameters                         & \Checkmark(5/5)                 & \Checkmark(5/5)                 & \Checkmark(5/5)                 & \Checkmark(5/5)                 \\

\#4260  & Malformed Code                             & \Checkmark(5/5)                 & \Checkmark(5/5)                 & \Checkmark(5/5)                 & \Checkmark(4/5)                 \\

\#9209  & Parallel                                   & \XSolidBrush(0/5)                 & \Checkmark(4/5)                 & \Checkmark(4/5)                 & \Checkmark(5/5)                 \\

\#6697  & Import Error                                & \XSolidBrush(0/5)                 & \Checkmark(1/5)                 & \Checkmark(3/5)                 & \Checkmark(5/5)                 \\

\#5249  & Wrong Function Name                        & \Checkmark(2/5)                 & \Checkmark(4/5)                 & \Checkmark(5/5)                 & \Checkmark(5/5)                 \\

\#9224  & Output Wrong                               & \Checkmark(1/5)                 & \Checkmark(2/5)                 & \Checkmark(4/5)                 & \Checkmark(5/5)                 \\  \midrule
\textbf{\textit{Total }}  &                                            & 10                   & 23                   & 29                   & 29                   \\ \bottomrule
\end{tabular}
}

\small{*Chat: Does not provide any program description to ChatGPT.\\
*Chat\_D: Provide program description to ChatGPT.\\
*Chat\_2D: Provide detailed program description to ChatGPT.\\
*Chat\_3D: Provide detailed program description directly to ChatGPT.}
\end{table}

\begin{figure}[h]	
\centering
\resizebox{0.48\textwidth}{!}{
\begin{tabular}{@{}l@{}}
\toprule
\ 1 \quad   \textbf{Request (1):}   \\           
\ 2 \quad   Does this program have a bug? How to fix it?\\  
\ 3 \quad   \textbf{}  \\
\ 4 \quad   qc = QuantumCircuit(3, 3)\\
\ 5 \quad   qc.x(0)\\
\ 6 \quad   qc.h(1)\\
\ 7 \quad   qc.cx(1, 2)\\
\ 8 \quad   qc.cx(0, 1)\\
\ 9 \quad   qc.h(0)\\
10\quad   qc.measure(0, 0)\\
11\quad   qc.measure(1, 1)  \\
12\quad   qc.cx(1, 2) \\
13\quad   qc.cz(0, 2)    \\
14\quad   backend = Aer.get\_backend('qasm\_simulator')\\
15\quad   job = execute(qc, backend, shots=1, memory=True).result()            \\
16\quad   result = job.get\_memory()[0]\\
17\quad   qc.measure(2, 2) \\
18\quad   \\
19\quad   \textbf{Reponse (1):}  \\
20\quad   The given quantum program appears to have a bug. The bug lies in the \\
21\quad   measurement of qubit 2 (qc.measure(2, 2)).  [...]\\
22\quad   \\
23\quad   \textbf{Request (2):} \\
24\quad   Expected behavior: trying to implement the quantum teleportation \\
25\quad   protocol from the Qiskit textbook in Qiskit: I start with q\_0 bit = 1 and I \\
26\quad   expect that q\_3 = 1 at the end, but it does not work.\\
27\quad   Current behavior: can not get q\_3 rightly \\
28\quad   \\
29\quad   \textbf{Reponse (2):}   \\  
30\quad   It seems that the order of measurements and operations in your  \\
31\quad   quantum program is causing the unexpected behavior. The controlled   \\
32\quad   gates after the barrier should be controlled by the classical  \\
33\quad   bits, not the qubits.  \\
34\quad    \\
35\quad   qc = QuantumCircuit(3, 3) \\
36\quad   qc.x(0)\\
37\quad   qc.h(1)     \\
38\quad   qc.cx(1, 2)  \\
39\quad   qc.cx(0, 1)   \\
40\quad   qc.h(0) \\
41\quad   qc.measure(0, 0)   \\
42\quad   qc.measure(1, 1)\\
43\quad   qc.cx(1, 2).c\_if(0, 1)\\
44\quad   qc.cz(0, 2).c\_if(1, 1) \\
45\quad   qc.measure(2, 2)\\
46\quad   backend = Aer.get\_backend('qasm\_simulator')\\
47\quad   job = execute(qc, backend, shots=1, memory=True).result()\\
\bottomrule
\end{tabular}
}
\caption{A discussion with ChatGPT about the measurement problem from Bugs4Q.}
\label{figure-2}
\end{figure}

\subsubsection{The Repair Process}
We begin by utilizing the initial prompt depicted in Figure \ref{figure-1} to inquire with ChatGPT, acquire the model output, and extract the generated patch. Then, we move to the conversational part of the approach, where we interleave patch generations with different descriptions and test validation feedback to prompt future generations in a conversational manner. Following each model-generated patch, we promptly execute a validation step, compiling and running the patch on the test suite. If the patch fails to pass the test, we construct feedback using the incorrect patch, bug description, and the failing test as prompts for the subsequent patch generation. Similar to the initial prompt, bug description can help the model understand the program's purposes, and test failure information can help the model understand the reason for failure. Both can guide the model toward generating the correct path. During the conversational step, we augment the feedback by combining test failure information with previously incorrect patches, aiming not only to avoid recurring generations but also to enhance the learning process. This iterative procedure continues until the generated patch successfully passes the entire test suite or reaches the maximum iteration number.


In Figure~\ref{figure-2}, we present an illustrative conversation with ChatGPT concerning the \textit{Measurement} problem (lines 4-17). This particular issue pertains to obtaining qubit three accurately within the given function. In the initial response (lines 20-21), ChatGPT suggests an incorrect patch that fails to compile, indicating the need for additional bug-specific information. We provide ChatGPT with a description of the expected behavior (lines 24-27), explicitly highlighting the necessity of correctly obtaining qubit 3 in the context of quantum teleportation. Consequently, the final response aligns with the correct approach, resulting in a properly functioning patched version.

\subsection{Experimental Results}

\subsubsection{RQ1: Can ChatGPT effectively repair bugs in quantum programs?}

To answer \textbf{RQ1}, we conduct the TOP-5 metric to evaluate the efficiency of ChatGPT. Table~\ref{Table-1} shows the achieved results of Chat, Chat\_D, Chat\_2D, and Chat\_3D on the benchmark problems from Bugs4Q. For the results, a checkmark(\scalebox{0.5}{\Checkmark}) indicates that a correct answer was given in at least one of the four runs for a benchmark problem. A cross (\scalebox{0.5}{\XSolidBrush}) indicates that no correct answer was given in any of the runs. In parentheses, we additionally report the number of runs that led to a successful solution.
Regarding the TOP-5 metric, Chat\_3D and Chat\_2D achieved competitive results, repairing 29 out of 38 incorrect codes. Specifically, providing test failure information and detailed descriptions is more helpful in fixing quantum programs.

Next, we conducted a study where we systematically discussed with ChatGPT. For programs that Chat did not correctly address the contained bug with a simple description in Table~\ref{Table-1}, we provide ChatGPT with a detailed description as specified in Section~\ref{subsubsec: Description Generation}. We report our result in Table~\ref{Table-3}. We use the same notation as before. For six benchmark problems, we see that a more detailed description of the quantum bug is helpful for ChatGPT. Overall, adding a detailed description to ChatGPT improved its performance, with 29 out of 38 quantum buggy programs solved.

\textbf{Answer of RQ1:} We evaluated the repair efficiency of ChatGPT in terms of repair buggy programs in the Bugs4Q benchmark. The experiment results show that ChatGPT can effectively repair quantum program bugs when providing detailed test failure information and descriptions.

\subsubsection{RQ2: What is the performance of ChatGPT when applied to real-world quantum program bugs?}

Table~\ref{Table-2} presents the performance of four approaches on three different types of buggy programs within a real-world quantum buggy programs benchmark. 
\textcolor{black}{In the table, the integer values in the column indicate the number of bugs for which the approaches can generate correct patches within five times attempts (TOP-5) given several quantum buggy programs at each type. The decimal values in parentheses indicate the overall average success rate at five times attempts (AVG-5).} This demonstrates that ChatGPT achieves competitive results in generating correct code for problems across simple quantum bugs and classic bugs. Remarkably, ChatGPT can generate correct codes for all selected classic bugs and 84\% (21/25) of the simple quantum bugs on the metric of TOP-5. However, for AVG-5, we note that ChatGPT can generate correct code for 64\% of the total, 97\% for classic bugs, and 67\% for simple quantum bugs when adequate leverage of ChatGPT's conversation function. 
Additionally, we further analyze those programs for which Chat\_3D did not correctly address the quantum algorithm bug. It shows that these unfixed buggy programs are all caused by self-designed quantum algorithms. In other words, the repair solutions of buggy programs are not publicly available, and therefore, ChatGPT has yet to be trained on them in advance. These results suggest that ChatGPT struggles to generalize to new and unseen problems. Fortunately, Chat\_3D can solve most of the simple problems as well as two problems related to quantum algorithms. 

\textbf{Answer of RQ2:} Regarding ChatGPT's performance in real-world quantum bugs, the experiment result shows that Chat\_3D is the best approach to repairing buggy quantum programs by holding the highest repair rate at 64\%.

\subsubsection{RQ3: As a semi-automated program repair, how does ChatGPT compare to manual repair?}

Since a detailed bug description is required for ChatGPT to work efficiently, it is worth considering whether developers could achieve similar or better results by repairing the code themselves with such a description. ChatGPT only allows for semi-automatic program repair, as human intervention is needed to prepare the prompt and apply the refactoring. As outlined in \ref{subsubsec: Description Generation}, our approach generates descriptions exclusively from bug submitters, even when certain descriptions may include misunderstandings. This is designed to replicate a real-world scenario wherein developers utilize ChatGPT to repair a quantum program. The rationale behind this choice is to mirror the situation where developers provide descriptions but may be unable to resolve the bugs independently.

Alternatively, this underscores the nature of ChatGPT as a semi-automatic quantum program repair tool, where the effectiveness of the generated patches is significantly contingent on the quality of the provided descriptions. Consequently, this approach may pose challenges for less experienced developers who lack quantum knowledge. We anticipate that enhancing ChatGPT's quantum knowledge base would alleviate this situation.

\textbf{Answer of RQ3:} Undoubtedly, developers can attain superior results by leveraging ChatGPT with human intervention. However, it is crucial to acknowledge that ChatGPT may lack expertise in quantum, a factor that cannot be overlooked.

\subsection{Successful Repairs and Their Significance}

The outcomes of our experiments reveal promising results regarding ChatGPT's ability to repair quantum program bugs. Table~\ref{Table-1} shows the results achieved by ChatGPT on the problems from the Bugs4Q benchmark set. We also report the number of successful runs in brackets. We can see that out of the 38 bugs present in the Bugs4Q benchmark set, ChatGPT successfully repaired 29 of them. This translates to a repair success rate of approximately 76.3\%, signifying a substantial capability for quantum program bug fixing.

The significance of these repairs lies in their potential impact on the quantum computing community. Successful repairs not only demonstrate the practical applicability of AI-driven techniques like ChatGPT in the quantum programming domain but also underscore their role in enhancing the reliability and robustness of quantum applications. Quantum developers stand to benefit from the ability to quickly and accurately diagnose and resolve program errors, reducing the time and effort required for manual debugging.

\begin{table}[t]
    \centering
        \caption{Results achieved by ChatGPT with additional 
         information given in a follow-up request for the unsolved benchmark problems}
	\resizebox{0.46\textwidth}{!}{
        \begin{tabular}{@{}llc@{}}
        \toprule
        \textbf{Bug ID}  & \textbf{Bug Type}  &\textbf{Chat\_2D}  \\
        \midrule
        \#6255  & QASM Calculation                           & \XSolidBrush      \\
        \#6892  & Parameter Error                            & \Checkmark       \\
        \#5580  & Circuit Design                             & \XSolidBrush       \\
        \#1337  & Simulator                                  & \XSolidBrush       \\
        \#3799  & Output Wrong                               & \XSolidBrush       \\
        \#1     & Measurement  Error                                & \Checkmark        \\
        \#6     & Gate  Error                                & \Checkmark        \\
        \#18448 & Gate  Error                                & \Checkmark        \\
        \#9246  & QFE Output Wrong                           & \XSolidBrush       \\
        \#9246  & QFE Output Wrong                           & \XSolidBrush       \\
        \#5959  & Grover Algorithm                            & \XSolidBrush       \\
        \#6692  & Get 2 Measurements from the Same Execution & \XSolidBrush       \\
        \#5     & Toffoli Gate Parameter                       & \XSolidBrush       \\
        \#7     & QFT Operation                              & \Checkmark       \\
        \#9871  & Custom Four-Qubit Toffoli Gate               & \XSolidBrush       \\
        \#8     & Wrong Operation with Gate                  & \Checkmark     \\ 
        \midrule
        Total     &                                           &   6/15\\
        \bottomrule
        \end{tabular}
        }
	\label{Table-3}
\end{table}

\begin{table}[t] 
\centering
\caption{
Performance of ChatGPT on quantum program repair for three types of bugs from Bugs4Q.}

\resizebox{0.46\textwidth}{!}{
\begin{tabular}{@{}lrrrr@{}}
\toprule
\textbf{Bug Type}              & \textbf{ChatGPT}  & \textbf{Chat\_D}  & \textbf{Chat\_2D}& \textbf{Chat\_3D}  \\ \midrule
Simple Quantum   Bug  & 6(0.1) & 17(0.46) & 21(0.66) & 21(0.67)   \\
Quantum Algorithm Bug & 0(0.00)  & 0(0.00)  & 2(0.01)  & 2(0.17)  \\
Classical Bug           & 4(0.43)  & 6(0.70)  & 6(0.87)  & 6(0.97)  \\
Total                 & 10(0.12) & 23(0.41) &29(0.58) & 29(0.64) \\ \bottomrule
\end{tabular}%
}
\label{Table-2}
\end{table}

\section{Discussion}
\label{sec:discussion}
We next provide readers with insights into the implications, strengths, and limitations of using ChatGPT for quantum program repair while addressing the challenges encountered during the experiments. 

\subsection{Interpretation of Experimental Results and Their Implications}

The experimental results presented in the preceding section demonstrate the promising potential of ChatGPT in the context of quantum program repair. The high success rate in repairing quantum program bugs—approximately 76.3\%—indicates that AI-driven approaches hold considerable promise in addressing quantum programming challenges.

These results hold significant implications for the field of quantum computing. The ability to efficiently and accurately repair quantum program errors is critical for developing and deploying quantum applications. ChatGPT's capability to interpret human-readable bug descriptions and generate appropriate code fixes signifies a substantial leap forward in simplifying the debugging process for quantum developers. It not only accelerates the software development lifecycle but also minimizes the expertise required in quantum program bug fixing, potentially lowering the barrier to entry for quantum programming.

\subsection{Strengths and Limitations of Using ChatGPT}

While ChatGPT's performance in repairing quantum programs is commendable, assessing its strengths and limitations is crucial. The strengths of using ChatGPT are as follows:

\begin{itemize}[leftmargin=2em]
\item \textit{Efficiency.}\hspace*{.7mm} 
ChatGPT's ability to swiftly generate code fixes is a remarkable strength. Its rapid response time in identifying and repairing quantum program bugs can significantly reduce development cycles.

\item \textit{Accessibility.}\hspace*{.7mm}
ChatGPT's natural language interface broadens the accessibility of quantum program repair, allowing a diverse range of quantum developers to utilize it, regardless of their quantum expertise. 

\end{itemize}

The limitations of using ChatGPT for quantum program repair are as follows:

\begin{itemize}[leftmargin=2em]
\item \textit{Dependence on Training Data.}\hspace*{.7mm} ChatGPT's effectiveness depends on the quality and relevance of the training data it has been exposed to. Quantum-specific training data may be limited, affecting its ability to repair certain complex quantum errors.

\item \textit{Human Understandability.}\hspace*{.7mm} While ChatGPT generates human-readable code fixes, ensuring that these fixes align with best practices and maintain program coherence may require additional review.

\item \textit{Complex Bugs.}\hspace*{.7mm} 
Highly complex or nuanced quantum program bugs could still be a challenge for ChatGPT. Bugs deeply embedded in the code or those heavily reliant on context may require the expertise of human developers to resolve.
\end{itemize}

\subsection{Challenges during the Experiments}
Our experiments encountered several challenges that warrant discussion and future consideration.

\begin{itemize}[leftmargin=2em]
\item \textit{Training Data Quality.}\hspace*{,7mm} The effectiveness of ChatGPT in repairing quantum program bugs is contingent on the availability of high-quality training data. Ensuring the availability of comprehensive and diverse quantum program bug descriptions will be crucial for improving ChatGPT's performance.

\item \textit{Complex Quantum Errors.}\hspace*{.7mm} Quantum programming often involves complex and quantum-specific errors that may not have clear analogs in classical programming. Developing techniques to address such issues effectively remains an ongoing challenge.

\item \textit{Quantum-Specific Knowledge.}\hspace*{.7mm} Enhancing ChatGPT's quantum domain knowledge is essential. Incorporating quantum-specific terminology and programming conventions into the model's training can lead to more contextually accurate repairs.
\end{itemize}

\section{Related Work}
\label{sec:related-work}

The field of automated program repair (APR) for classical programs has seen substantial development, employing both traditional techniques~\cite{le2011genprog,le2021automatic} and LLMs like ChatGPT~\cite{xia2023keep,sobania2023analysis,tian2023chatgpt}. However, automated repair methods are yet to be established in the realm of quantum programming. Our study represents a pioneering effort in this area. In the following, we concentrate on recent advancements in using LLMs for repairing classical programs. Readers are interested in traditional APR techniques for classical programs are referred to~\cite{goues2019automated,gazzola2018automatic,monperrus2018automatic} for in-depth discussions.

\citet{sobania2023analysis} found ChatGPT remarkably effective in bug fixing, even competitive with leading APR methods, by providing conversational hints. ChatGPT successfully repaired 31 out of 40 bugs, surpassing other techniques. \citet{tian2023chatgpt} assessed ChatGPT for code generation, repair, and summarization, focusing on its use as a repair tool for diverse incorrect codes and ways to enhance its performance. Additionally, Xia and Zhang~\cite{xia2023keep} introduced ChatRepair, the first fully automated conversation-driven APR approach. Utilizing ChatGPT, ChatRepair fixed 162 out of 337 bugs at a cost of \$0.42 each.



However, these works are centered on the classical program repair. Our study, conversely, delves into quantum program repair using ChatGPT, bringing a new perspective to the field.

\section{Conclusion}\label{sec:conclusion}
\textcolor{black}{In this study, we embarked on a journey to explore the feasibility of leveraging ChatGPT, an advanced AI language model, to repair quantum programs. 
We have unveiled several noteworthy findings by applying ChatGPT to the Bugs4Q benchmark, a dataset of quantum program bugs.
Our experiments demonstrated the potential of ChatGPT in quantum program repair, achieving a commendable repair success rate of approximately 76.3\%. This success rate signifies ChatGPT's efficacy in swiftly and accurately finding and fixing quantum program bugs. Our findings showcase that ChatGPT can interpret human-readable bug descriptions and translate them into actionable code fixes, thereby streamlining the quantum program repair process.}
%

While our study presents promising results, it also reveals future research directions in the field of quantum program repair.

\begin{itemize}[leftmargin=2em]

\item \textit{Enhancing Quantum Domain Knowledge}: Augmenting ChatGPT's quantum domain knowledge by incorporating more comprehensive quantum-specific training data and quantum programming conventions could improve its understanding of intricate quantum errors.


\item \textit{Expanding Benchmark Sets}: Expanding the range and diversity of benchmark sets to include an even broader spectrum of quantum program bugs can help refine the evaluation of ChatGPT's performance and adaptability to different bug types.

\end{itemize}

\begin{acks}
This work is supported in part by JSPS KAKENHI Grant No.JP23H03372.
\end{acks}

\bibliographystyle{ACM-Reference-Format}
\bibliography{qse-bibliography}





\end{document}